\documentclass[
amsmath,
amssymb,
aps,
superscriptaddress,
twocolumn]{revtex4-2}

\usepackage[utf8]{inputenc}

\usepackage{graphicx}
\usepackage{amsmath}
\newcommand{\fig}{Fig.~}

\newcommand{\uni}{Department of Computer Sciences, University of T\"ubingen, Germany}
\newcommand{\mpi}{Max Planck Institute for Biological Cybernetics, T\"ubingen, Germany}

\begin{document}
\title{Network bottlenecks and task structure control the evolution of interpretable learning rules in a foraging agent}

\author{Emmanouil Giannakakis} 
\affiliation{\uni}
\affiliation{\mpi}

\author{Sina Khajehabdollahi}
\affiliation{\uni}

\author{Anna Levina}
\affiliation{\uni}
\affiliation{\mpi}

\begin{abstract}
Developing reliable mechanisms for continuous local learning is a central challenge faced by biological and artificial systems. Yet, how the environmental factors and structural constraints on the learning network influence the optimal plasticity mechanisms remains obscure even for simple settings. 
To elucidate these dependencies, we study meta-learning via evolutionary optimization of simple reward-modulated plasticity rules in embodied agents solving a foraging task. 
We show that unconstrained meta-learning leads to the emergence of diverse plasticity rules. However, regularization and bottlenecks to the model help reduce this variability, resulting in interpretable rules. 
Our findings indicate that the meta-learning of plasticity rules is very sensitive to various parameters, with this sensitivity possibly reflected in the learning rules found in biological networks. When included in models, these dependencies can be used to discover potential objective functions and details of biological learning via comparisons with experimental observations.

\end{abstract}

\maketitle

\section{Introduction}
A hallmark of living organisms is their ability to adapt to their environment and assimilate new information to modify their behaviour. It is unclear how the ability to learn first evolved \citep{Papini2012}, yet their advantages are clear. Natural environments are too complex for all the necessary information to be encoded genetically \citep{SnellRood2013} and more importantly, they keep changing during an organism’s lifetime in ways that cannot be anticipated \citep{Ellefsen2014, Dunlap2016}. The capacity to learn is so beneficial in evolutionary terms that most organisms accept high costs, such as increased energy consumption \citep{Mery2005} and the need for lengthy nurturing periods \citep{Thornton2011} in order to maintain an ability to learn throughout their lifetimes.

The most prominent mechanism by which learning occurs in biological organisms is via changes to the strengths of synaptic connections between neurons  \citep{Citri2008, Feldman2009, Wickliffe2019}. Synaptic plasticity is observed across a wide variety of organisms and remains active throughout most organisms' lifetimes \citep{Power2016}. Starting from the early days of computational neuroscience \citep{he1948organisation, BCM1982}, many experimental and theoretical studies have focused on identifying the basic principles of synaptic modification and developing simplified models of these processes \citep{Abarbanel2002, Hennig2013, Kumarapathirana2021, Magee2020}. 

Using an automated discovery of optimal plasticity rules for solving particular tasks, \emph{meta-learning of plasticity}, can be successfully applied for training artificial neural networks. There the advantage of plasticity lies in its time-continuous nature (compatible with life-long learning) and its local nature \citep{Bengio1991, Soltoggio2018, Iwhiwhu2020}.
Often this meta-learning is implemented using evolutionary computations \citep{Najarro2020, Pedersen2021, Yaman2021}. 
Several applications following this approach can compete with state-of-the-art machine learning algorithms on various complex tasks \citep{Burms2015, metz2018learning, lindsey2019learning}.

The meta-learning of synaptic plasticity rules in artificial networks is a promising approach for investigating the functions of synaptic plasticity in biological networks. Experimental measurements of synaptic plasticity can correlate the impact that pre- and post-synaptic activity have on the synaptic strength but can rarely identify the objective function that the learning process is optimising for or the exact impact different plasticity parameters have on the learning task \citep{Richards2019}. Optimizing learning rules in artificial networks can help uncover the functions of different plasticity forms in biological networks \citep{Confavreux2020, Jordan2021, Tyulmankov2022, Shervani2023, confavreux2023metalearning, Mehta2023}. 

Still, most of the meta-learning studies focus on the inference of the optimal rules for minimising specific objective functions while ignoring other biophysical constraints that may influence the learning process. In particular, while it has been shown that network structure \citep{Giannakakis2023} can excerpt a strong effect on the performance of synaptic plasticity and that multiple compensatory processes heavily interact with classical plasticity mechanisms is biological systems \citep{Zenke2016}, such factors are rarely considered in meta-learning studies of plasticity. Additionally, a recent study  \citep{Ramesh2023} has demonstrated that meta-learning can uncover diverse families of plasticity rules leading to identical learning outcomes. While this finding encourages a more holistic understanding of plasticity that focuses on shared learning outcomes, rather than individual rules, it nevertheless raises the need for methods to classify and assess learning rules in different settings.

Here, extending previous work \citep{Giannakakis2022}, we model embodied agents equipped with plastic sensory networks that learn about the agents environment during a foraging task. We study the effect that different structural features of the network and parameters of the evolutionary process have on the form of evolved functional reward-modulated plasticity rules. We find that the apparent redundancy of evolved learning rules can be significantly reduced by the introduction of simple regularization techniques and information bottlenecks. We further find that small changes in the plastic network's structure (such as changes in the activation function of a neuron or the method of weight normalization) can have a significant effect on the form of the meta-learned plasticity rules. Finally, we separate the plastic sensory networks from the moving agents and we attempt to re-create the plasticity rules that co-evolved with the agents motor networks. We find that different assumptions about the function of the motor networks as well as the function that the plasticity performs in the embodied system can lead to the emergence of distinct rules in the static sensory networks. We argue that such effects can enable us to infer important facts about the objective functions and learning processes of biological learning systems.

\section{Methods}
Our model consists of an environment with food particles of different values and agents that can sense the nearest food to them and decide what to eat. To this end, the agent must move to the target food using its motors and the motor network that controls them. The sensory system of the agents can learn using synaptic plasticity, which is encoded genetically and evolves together with the agent's motor network. In the following, we describe the parts of the model in more detail. 

\subsection{Environment}

A number of food particles (usually 100, unless otherwise specified) are randomly placed in a $300 \times 300$, 2D space with periodic boundary conditions. Each food particle consists of various amounts of $N$ ingredients with positive (food) or negative (poison) values.  The value of a food particle is a weighted sum of its ingredients. To predict the reward value of a given resource, the agent must learn the values of these ingredients by interacting with the environment.

The values of the $N$ ingredients $W^c=\{W^c_1,\ldots, W^c_N\}$, with $W^c_i = -1 + \frac{2(i-1)}{N-1}, \; i=1,\ldots, N$ are equally spaced between -1 and 1. 
Each food particle is encoded as an $N$-dimensional vector $X_t = (x_1, \dots, x_N)$ is presented, where the value $x_i, \ i \in \{1, \dots, N\}$ represents the quantity of the ingredient $i$. 
The value of the food particle is $W^c X^T = \sum_i W^c_i \cdot x_i$. 
We draw $x_i$ independently form a uniform distribution on the $[0,1]$ interval ($x_i \sim U(0, 1)$). Each time a food particle is eaten, it is re-spawned with the same value somewhere randomly on the grid (following the setup of \citep{Khajehabdollahi2022}). This re-spawning algorithm guarantees that even if the agents have already evolved to eat only non-poisonous food, the mean of the food value distribution in the environment remains $\approx 0$. 

\subsection{Agent}

\begin{figure*}[!htb]
\centering
\includegraphics[width=0.95\textwidth]{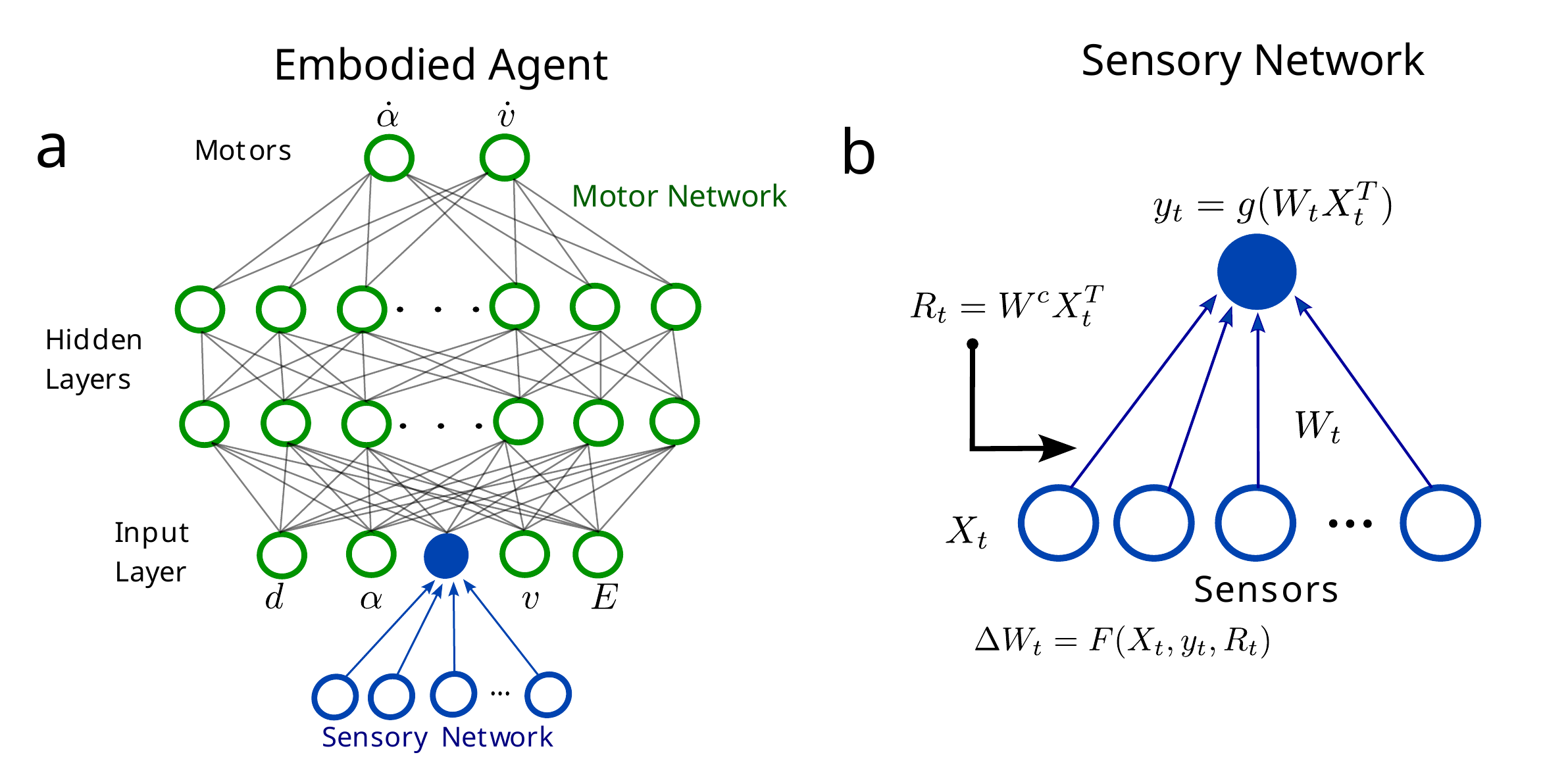}
\caption{\textit{Structure of the neural network controlling the embodied foraging agent. \textbf{a.} A diagram of the network controlling the foraging agent. The sensor layer receives inputs at each time step (the ingredients of the nearest food). The output of that network is given as input to the motor network, along with the distance $d$ and angle $\alpha$ to the nearest food, the current velocity  $v$, and energy $E$ of the agent. These signals are processed through two hidden layers to the final output of motor commands as the linear and angular acceleration of the agent \textbf{b.} Details of the sensory network. The sensor layer receives inputs representing the quantity of each ingredient of the nearest food at each time step. The agent outputs its assessment of the food's value $y_t$, and when a food particle is consumed it receives the true value $R_t$ as feedback; it finally uses this information to update the weight matrix according to the plasticity rule.}}
\label{fig:networks}
\end{figure*}

The agent (\fig\ref{fig:networks}a) consists of a motor network, which controls their movement across the 2D space and a plastic sensory network that learns the values of different ingredients via synaptic plasticity and provides the agent with information about the value (food or poison) of the nearest food (\fig\ref{fig:networks}b).

\subsubsection{Motor network}

An agent's motor network is not plastic but trained via evolution across generations and remains constant throughout an agent's lifetime. The motor network consists of 4 layers. An input layer with sensors that provide the agents with information such as the distance from the nearest food, the angle between the current velocity and the nearest food direction, its own velocity, and its own energy level (sum of consumed food values). Finally, one of the sensors is connected to the plastic sensory network, which provides the agent with an estimation of the nearest food's value. Two hidden layers (of 30 and 15 neurons with $\tanh$ activation) process these inputs. The network's outputs are the angular and linear acceleration of the agent.

\subsubsection{Sensory network}

The sensory network consists of $N$ sensory neurons (one for each type of the ingredients, loosely resembling the olfactory system of \emph{drosophila melanogaster} \citep{Aso2014, Fabian2023} ) that are projecting to a single post-synaptic neuron signalling the food value to the motor network.  At each time step, an input $X_t = (x_1, \dots, x_N)$, representing the nearest food to the agents, is given. The postsynaptic neuron outputs an assessment of the food $X_t$ value as $y_t = g(W_t X_t^T)$. Throughout the paper, $g$ will be either the identity function, in which case the prediction neuron is linear, or a step-function; however, it could be any other nonlinearity, such as a sigmoid or ReLU. When a food particle is consumed, the agent receives the real value of the input $R_t= W^c X_t^T$ as feedback for learning. Once an agent consumes a food particle, the input $X_t$, output  $y_t$ and feedback $R_t$ are used by the plasticity rule to update the sensory weights $W_{t+1} \leftarrow W_{t} + \Delta W_t$ where  $\Delta W_t = F(X_t, y_t, R_t)$.

\subsection{Plasticity rule parametrization} 
Reward-modulated plasticity is one of the most promising explanations for biological credit assignment \citep{Legenstein2008}. In our network, the plasticity rule that updates the weights of the sensor network is a reward-modulated rule which is parameterized as a linear combination of the input, the output, and the reward at each time step:
\begin{multline}
    \Delta W_t = F(X_t, y_t, R_t) = \eta_p [ R_t \cdot \overbrace{ (\theta_1 X_t y_t + \theta_2 y_t  +   \theta_3 X_t + \theta_4)}^{\text{Reward Modulated}} \\ +  \underbrace{(\theta_5 X_t y_t + \theta_6 y_t +   \theta_7 X_t  + \theta_8)}_{\text{Hebbian}} ].
\end{multline}
Additionally, after each plasticity step, the weights are normalized to maintain a constant sum, an important step for the stabilization of Hebbian-like plasticity rules \citep{Zenke2016}. Unless specified otherwise, the normalization happens via mean subtraction.

We use a genetic algorithm to optimize the learning rate $\eta_p$ and amplitudes of different terms $ {\theta} = (\theta_1, \dots, \theta_8)$. The successful plasticity rule after many food presentations must converge to a weight vector that predicts the correct food values (or allows the agent to correctly decide whether to eat a food or avoid it). 

To have comparable results, we divide $ {\theta} = (\theta_1, \dots, \theta_8)$ by  $\theta_\mathrm{max} = \max_{k}|\theta_k|$. So that ${\theta} / \theta_\mathrm{max} = {\theta}^\mathrm{norm} \in [-1,1]^8$. We then multiply the learning rate $\eta_p$ with $\theta_\mathrm{max}$ to maintain the rule's evolved form unchanged, $\eta_p^\mathrm{norm} = \eta_p \cdot \theta_\mathrm{max}$. In the following, we always use normalized $\eta_p$ and ${\theta}$, omitting $^\mathrm{norm}$.

\subsection{Evolutionary algorithm}
To evolve the plasticity rule and the moving agents' motor networks, we use a simple genetic algorithm with elitism \citep{Deb2011}. Each agent is initialized in a separate instance of the 2D environment. The agents' evolvable parameters (motor weights and plasticity parameters) are initialized at random (drawn from a Gaussian distribution), then the sensory network is trained by the plasticity rule and after 5000 time steps the agent is evaluated. The energy of an agent at the beginning of its lifetime is initialized to be positive ($E_0 = 2$) but if the agent's energy becomes negative during its lifetime (due to eating too many negative food particles), the agent can no longer move.
The fitness of an agent is its energy at the end of its lifetime, given as the sum of the values of the foods it consumed:
\begin{equation}
    E = \sum_{t \in T_{\text{eat}}}R_t
    \label{eq:fitness}
\end{equation}
where $T_{\text{eat}}$ are the timesteps where the agents consumed a food particle. After each generation, the best-performing agents (top 10 \% of the population size) are selected and copied into the next generation. The remaining 90 \% of the generation is repopulated with mutated copies of the best-performing agents. We mutate agents by adding independent Gaussian noise ($\sigma = 0.1$) to its parameters. Unless specified otherwise, we train a population of 100 agents for 250 generations.

\subsection{Curriculum}
The task of evolving the motor network weights and plasticity rule from scratch is very hard.  
To assist the evolutionary process of the moving agents, we developed a curriculum that progressively initializes the agents' sensory networks farther from the ground truth, making correct plasticity rule more and more essential to distinguish eatable from poisonous food particles rapidly. To do so, we add a normally distributed noise ($\mathcal{N}(0,\, \sigma)$) to the ground truth values with the progressively increasing variance per curriculum $\sigma \in \{0, 1, 2, 3, 4, 5\}$. The populations train within each curriculum for 250 generations before moving to the next curriculum step. All results presented are from populations trained until the final curriculum step.

\section{Summmary of previous results}

In previous work \citep{Giannakakis2022}, we studied the effect of different factors (environmental fluctuations, reliability of sensory input, and task complexity) on the evolved learning rate and form of reward-modulated plasticity rules. In particular, we investigated the speed at which agents need to learn about their environment (i.e., learning rate). Here, we extend that study by focusing on the impact of network and task parameters on the resulting plasticity rules.

\section{Results}

\begin{figure*}[!htb]
\begin{center}
 \includegraphics[width=0.99\textwidth]{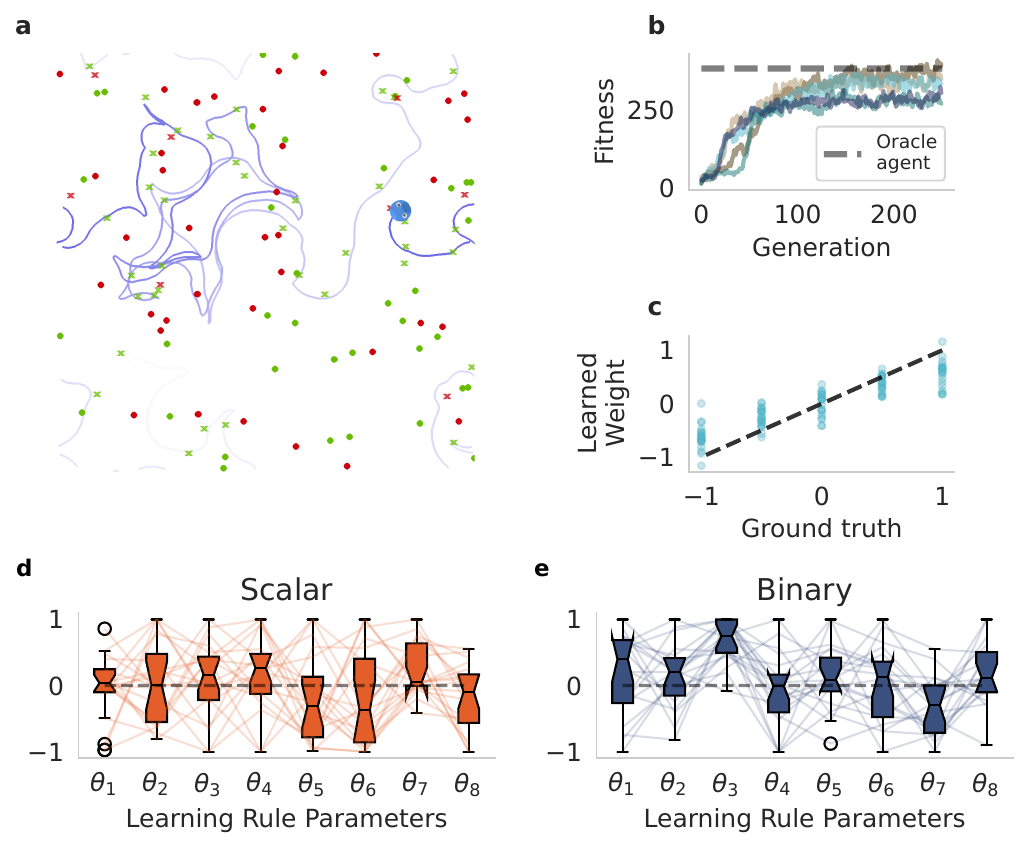}
\caption{Diverse learning rules lead to high performance in the foraging task \textit{\textbf{a.} The trajectory of an agent (grey line) in the 2D environment. A well-trained agent will approach and consume food with positive values (green dots) and avoid negative food (red dots). The $\times$ signs denote the locations of food particles consumed by the agent \textbf{b.} The  fitness of the best performing agent from 5 evolving populations (Eq.~\ref{eq:fitness}) increases over generations of the evolutionary algorithm. The dotted gray line indicates the maximum fitness of the "Oracle" non-plastic agents who are given the correct food values as input \textbf{c.} The learned weights of the sensory network (blue dots) correlate strongly with the actual ingredient values (Pearson cc of $0.92 \pm 0.09$). \textbf{d, e.} The evolved plasticity rules across 20 runs for agents with scalar/binary readout, respectively.}}
\label{fig:convergence}
\end{center}
\end{figure*}

 \subsection{The agents evolve to solve the foraging task}

 We begin by evolving 20 populations of 100 agents with a linear sensory readout ($g(x) = x$). At each step of the curriculum, the agents evolve a functional motor network and a plasticity rule for the sensory network. The converged sensory weights allow the agent's motor network to adequately interpret the quality of food, enabling it to effectively approach food particles with positive values while avoiding the ones with negative values (\fig\ref{fig:convergence}a).

 We see that after $\sim$ 100 evolutionary steps (\fig\ref{fig:convergence}b), the agents reach a stable high fitness ($> 250$). To put this value into context, we train 5 populations of "Oracle" agents without a plastic sensory network, who are given the real values of nearby foods as direct input. These populations of agents reach a maximum fitness of  $383 \pm 37$ at the end of their training. Thus, the plastic agents' fitness is about as good as they could possibly get. This effectively means that the best-performing agents can accurately navigate and rapidly learn the correct ingredient values via their plastic sensory networks and subsequently only consume foods with positive values.

 The weights of the best-performing agent's sensory networks subject to synaptic plasticity converge to the values strongly correlated (Pearson correlation coefficient of $0.92 \pm 0.09$) but not identical to the correct ingredient values (\fig\ref{fig:convergence}c). Thus, the evolutionary algorithm produces learning rules that can reliably learn a rough approximation of the real distribution of ingredient values, enabling the motor network to decide which food particles the agent should consume.

 \subsection{Redundancy in the plasticity rules and its diminution by the information bottleneck}

 Once the evolution has converged, we look at the evolved plasticity parameters $ {\theta} = (\theta_1, \dots, \theta_8)$ of the best-performing agents of each population (\fig\ref{fig:convergence}d). We observe that despite the fact that all the rules seem to be converging to similar sensory weights (\fig\ref{fig:convergence}c), the plasticity parameters are quite different between different rules and effectively present no visible pattern. Such redundancy of plasticity rules has been observed in previous studies \citep{Ramesh2023} of meta-learning.

We hypothesize that the redundancy of learning rules is due to the superfluous information the sensory network passes on to the motor network. For good task performance, the motor network only needs to know whether to approach or avoid a food particle (depending on whether its value is positive or negative). Since the output of the sensory network is a scalar value, any learning rule which produces output that allows the network to distinguish between positive and negative foods will perform equally well.

To test this hypothesis, we introduce a bottleneck of information propagation between the sensory and motor networks by using a step-function nonlinearity on the output of the sensory network:
\begin{equation}
g(x) =
    \begin{cases}
        1, & \text{if  } \ x \geq 0, \\
        -1, & \text{if  } \ x < 0. \\
    \end{cases} \label{eq:nonlineariry}
\end{equation}
After convergence, we observe that despite the variability of the resulting learning rules being high, a general pattern begins to emerge in the form of the rules. Specifically, the third parameter $\theta_3$, is consistently positive and close to 1(\fig\ref{fig:convergence}e). This means that the term $R_t \cdot X_t$, which correlates the presence or absence of ingredients with positive/negative rewards becomes prominent in the rule, indicating some form of Hebbian-like learning.

 \subsection{The information bottleneck improves performance and generalizability}

 \begin{figure*}[tb]
\centering
\includegraphics[width=0.95\textwidth]{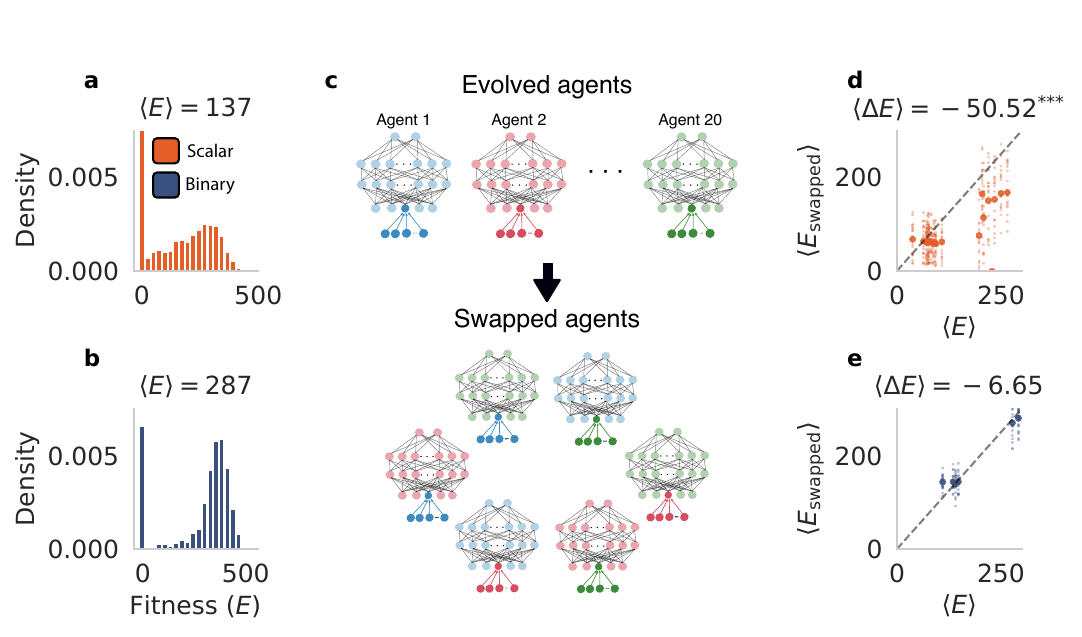}
\caption{\textit{Agents with binary sensory readouts perform and generalize better than those with scalar readouts.
\textbf{a, b}. The histogram of fitnesses of the top-ranking agent in each of the 20 runs over 100 independent environment realizations for scalar/binary networks times, respectively. \textbf{c}. Schematic of swapping networks: Motor and sensor networks are swapped between the fittest agents from each run. 
\textbf{d, e}. The swapped networks' fitnesses plotted against the mean fitness of the original configurations for the scalar/binary networks, respectively. The mean fitnesses for the swapped agents were significantly different (paired two-sided t-test) for the networks with a scalar readout ($p_{\text{value}} = 0.00087 $) but not for the ones with a binary readout ($p_{\text{value}} = 0.1457$).}}
\label{fig:Rules}
\end{figure*}

 We compare the performance of scalar and binary sensory readout agents in multiple realisations of the 2D environment. In both cases, we observe a high level of variability in the resulting finesses. 
 This variability is in part due to the fact that depending on the sequence of food particles an agent encounters early on during its lifetime, an agent can either learn the ingredient value distribution, which will lead to a relatively high fitness or encounter too many negative foods and remain stagnant during its lifetime (due to the energy constraint), which leads to a fitness of $\approx 0$ (\fig\ref{fig:Rules}b, d). This risk of an agent encountering too many negative foods at the beginning of its lifetime means that even very well-performing agents will fail for a small fraction of environment initializations. 
 Nevertheless, despite the high variability, we observe a significantly higher average fitness among the agents with binary (\fig\ref{fig:Rules}d, average fitness $\langle E \rangle = 287$), compared to the ones with scalar sensory readouts (\fig\ref{fig:Rules}b, average fitness $\langle E \rangle = 137$), suggesting that the information bottleneck helps with task performance.

 We then examine how well the evolved rules can generalize between different motor networks. We swap the sensory and motor networks of different evolved agents and test the fitness of the resulting mixed agents (\fig\ref{fig:Rules}a). We observe a statistically significant reduction ($p_{\text{value}} = 0.00087 $) in the average fitness (\fig\ref{fig:Rules}c) of agents re-combined in this way compared to agents with co-evolved sensory and motor networks in the case of scalar sensory readouts (here and later we use paired t-test comparing the mean fitness across 100 environment realizations for each tested agent against the mean fitness of the swapped agents with the same motor network - mean across 100 environments for each swapped agenda and 19 different swapped versions - that is approximately Gaussian). This suggests that the motor network evolves to interpret the specific output of a given learning rule, and thus, the differences between the converged sensory weights of different rules (\fig\ref{fig:convergence}c) make a significant difference in the resulting sensory readout.
 On the contrary, in the case of binary readouts (\fig\ref{fig:Rules}f), the average fitness remains the same, i.e. agents perform equally well with learning rules that co-evolved with the motor networks of different agents ($p_{\text{value}} = 0.1457$). Thus, adding the information bottleneck in the sensory network, not only improves performance, but it makes the agents much more generalizable and enables modular evolution

\subsection{Regularization of the plasticity parameters leads to interpretable rules} \label{Sec:Regularization}

\begin{figure*}[]
\centering
\includegraphics[width=0.90\textwidth]{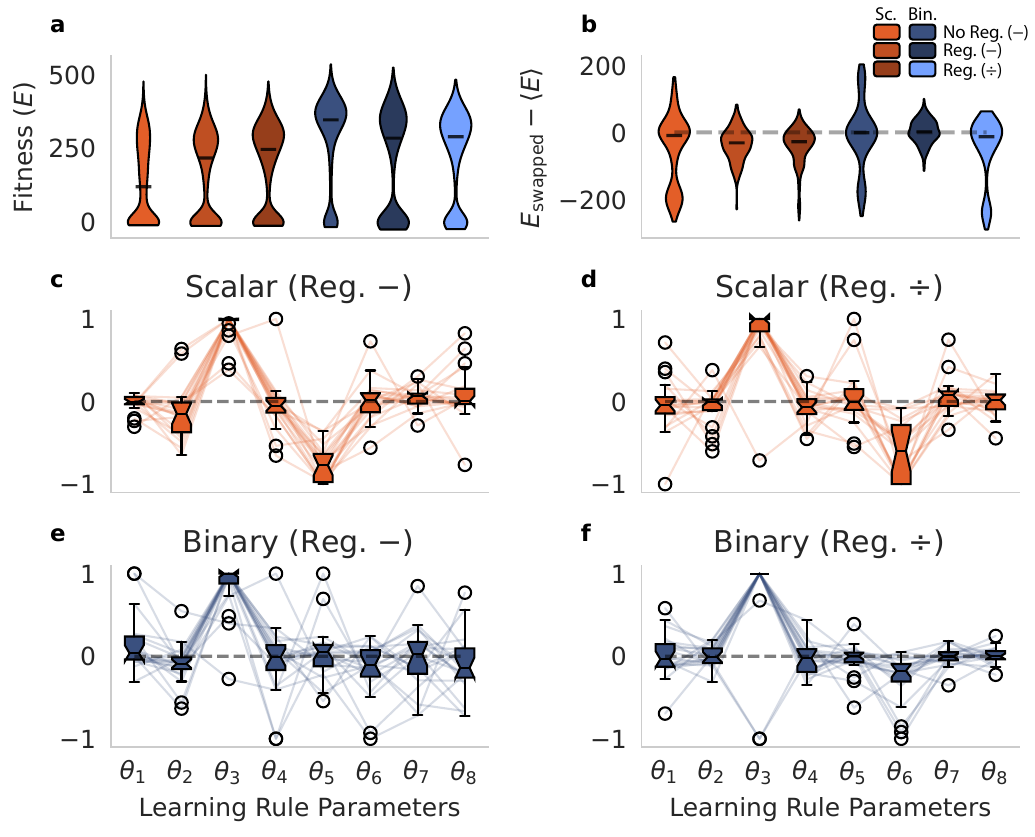}
\caption{\textit{ Plasticity rules converge with regularization. \textbf{a}. The fitness distribution for different simulations of scalar (Sc) and binary (Bin) sensory readouts, regularization of the plasticity parameters (Reg.) and subtractive (-) vs divisive (÷) weight normalization. \textbf{b}. The difference in fitness between the original and swapped agents for different simulations. \textbf{c}.The regularized rules developing in networks with scalar sensory readout for subtractive normalization. \textbf{d}. Same for divisive normalization. \textbf{e}. The regularized rules developing in networks with binary sensory readout for subtractive normalization. \textbf{f}. Same for divisive normalization. 
}}
\label{fig:Rules_L1}
\end{figure*}

We now test how the regularization of the plasticity rules decreases the outcome's variability. 
To this end, we regularize the plasticity parameters $\theta = \{\theta_1, \dots, \theta_8 \}$ using an $L^1$ norm, implemented as a weight decay of these parameters after each generation: $\theta_i \leftarrow \theta_i - \kappa \cdot \mathit{sign}(\theta_i)$, where  $\kappa$ is a weight-decay velocity taken in further experiments to be $\kappa = 0.05$. We evolve agents with the scalar and binary versions of the sensory network, and we observe that evolved plasticity rules show minimal rule patterns, different between binary and scalar networks  (\fig\ref{fig:Rules_L1}c, e). 

In the case of agents with a scalar sensory readout, the plasticity parameters  ${\theta} = (\theta_1, \dots, \theta_8)$  converge on approximately the same point for most of the evolved agents (\fig\ref{fig:Rules_L1}c). In particular, $\theta_3 \to 1$, $\theta_5 \to -1$, $\theta_i \to 0$ for all other $i$, and thus the learning rule converges to:
\begin{equation}
    \Delta W_t  = \eta_p [\theta_3 X_tR_t + \theta_5 X_ty_t] \approx \eta_p X_t (R_t - y_t).
\end{equation}
Since by definition $g(x) = x$ in this experiment, thus $y_t = g(W_t X_t^T) = W_t X_t^T$ and $R_t = W^c X_t^T$. For the weight update we get: 
\begin{equation}
     \Delta W_t = \eta_p X_t(W^c - W_t)X_t^T
\end{equation}
Thus, $\Delta W_t \rightarrow 0$ as $W \rightarrow W^c$, and consequently, this learning rule will match the agent's weight vector with the vector of ingredient values in the environment. We further observe a slight increase in the scalar agent's average fitness (\fig\ref{fig:Rules_L1}a), compare first and second violins).
Finally, we note that when mixing different scalar agent's sensory and motor networks, as in the previous section, the same reduction in the resulting fitness we observed for non-normalized rules persists. This suggests that despite the overall pattern being the same, differences between learning rules are still significant enough to require specialization of the motor networks for particular rules.

In the case of a binary readout, we see that all parameters converge to $\approx 0$ except for $\theta_3$ that converges to 1 (\fig\ref{fig:Rules_L1}e). This leads to a learning rule of the form:
\begin{equation}
    \Delta W_t  = \eta_p \theta_3 X_t \cdot R_t  \approx  \eta_p R_t \cdot (x_1, \dots, x_n)
\end{equation}
This rule clearly does not depend on the network's sensory weights $W_t$ and thus has no fixed points. Still, we see that the rule creates a Hebbian-like learning process: if a $w_i > 0 $ and the reward is positive $R_t>0$, then the absolute value of $w_i$ would increase; the same if $w_i < 0$ and  $R_t<0$, then the absolute value of $w_i$ would also increase. Thus the rule strengthens the weights that are positively correlated with positive rewards (and also increases the absolute value of the weights negatively correlated with negative rewards)
Overall, the learning rule leads to a sensory output $y_t$ that is highly correlated with the sign of the reward $R_t$ but develops weights with diverging absolute values.

For the binary sensory readouts, similar to the experiment in the previous section, swapping motor and sensory networks of different agents does not lead to a significant reduction in fitness, which indicates that binary readout networks maintain their generalizability and modularity under regularization of the plasticity parameters.

\subsection{The method of weight normalization impacts the evolved learning rule}

Most studies of Hebbian-like plasticity include a weight normalization mechanism that is necessary for stabilizing the learning dynamics over time \citep{Elliott2003, Zenke2016}.
However, a particular form of normalization can be essential for the learning outcome.     
As we have established in the previous sections, introducing an information bottleneck between the sensory and motor networks leads to superior performance and generalizability. Still, we showed that the resulting learning rule generates divergent weights. We hypothesize that this can be changed by adjusting the weight normalization.

So far, for normalizing the weights, we have used a mean subtraction after each plasticity step \citep{Miller1994, Goodhill1994}, which maintains the mean weight at $0$: 
\begin{align}
\begin{split}
W_t &= W_{t-1} + \Delta W_t \\
W_t &\leftarrow W_t - \langle W_t \rangle
\end{split}
\end{align}
However, it is just one option out of a wide range of mechanisms for normalizing synaptic weights that have been proposed in neuroscientific and AI studies \citep{Shen2021}. Moreover, some normalization mechanisms have been shown to have a significant effect on the performance of Hebbian learning \citep{Eckmann2022}. 
To examine how the choice of weight normalization mechanisms may affect the evolution of the learning rule, we repeated the experiments described in the previous section for both network settings (scalar and binary) using a divisive normalization.
This mechanism maintains the sum of the absolute values of the weights constant and equal to a given target $S_{g}$ which is set to $S_{g} = 3$ for all the experiments:
\begin{align}
\begin{split}
W_t &= W_{t-1} + \Delta W_t \\
W_t &\leftarrow W_t \cdot \frac{S_{g}}{\sum_{i = 1}^N |w_i(t)|},
\end{split}
\end{align}
Using this weight normalization mechanism after each plasticity step, we evolve 20 populations each, for both scalar and binary sensory networks. We see that the networks with a scalar sensory readout converge to a different learning rule (\fig\ref{fig:Rules_L1}d) than the equivalent networks with subtractive normalization (\fig\ref{fig:Rules_L1}c). Specifically, we see that instead of $\theta_5 \rightarrow -1$ and $\theta_6 \rightarrow 0$, under the divisive normalization $\theta_5 \rightarrow 0$ and $\theta_6 < 0$, which makes the learning rule to take the form:
\begin{equation}
     \Delta W_t = \eta_p [\theta_1 X_t R_t  + \theta_6 y_t]
\end{equation}
where $\theta_1 \approx 1$ and $\theta_6 < 0$ . In combination with the divisive normalization, this rule converges to a close approximation of the correct ingredient distribution $W^c$.

The trained agents maintain a similar fitness to previous experiments  (\fig\ref{fig:Rules_L1}b), and as expected from networks with scalar sensory readouts, the performance declines significantly ($p_{\text{value}} = 0.0003$, paired two-sided t-test) when we test agents with motor and sensory networks that did not co-evolve.

The networks with a binary sensory readout evolve a learning rule similar to all other binary readout networks (i.e. most parameters converge to the vicinity of $0$ except for $\theta_3 \rightarrow 1$). A small but significant difference between the networks that evolved with divisive normalization and those that evolved with subtractive normalization is that the former evolves a parameter $\theta_6 < 0$  (\fig\ref{fig:Rules_L1}f). The same pattern is observed much more prominently for networks with a scalar readout  (\fig\ref{fig:Rules_L1}d), which suggests that a negative $\theta_6$ parameter is important for networks with divisive normalization.

The divisive normalization successfully constrains the learned weights to be rather small (in contrast to other binary networks whose plasticity converges to very large sensory weights), and this does not seem to affect their fitness that remains relatively high (\fig\ref{fig:Rules_L1}a). Also, as with the other binary readout networks, swapping motor and sensory networks between different agents does not lead to a significant reduction in performance ($p_{\text{value}} = 0.067$, paired two-sided t-test).

\subsection{Trainable nonlinearity on the sensory readout}

\begin{figure*}[tb]
\centering
\includegraphics[width=0.99\textwidth]{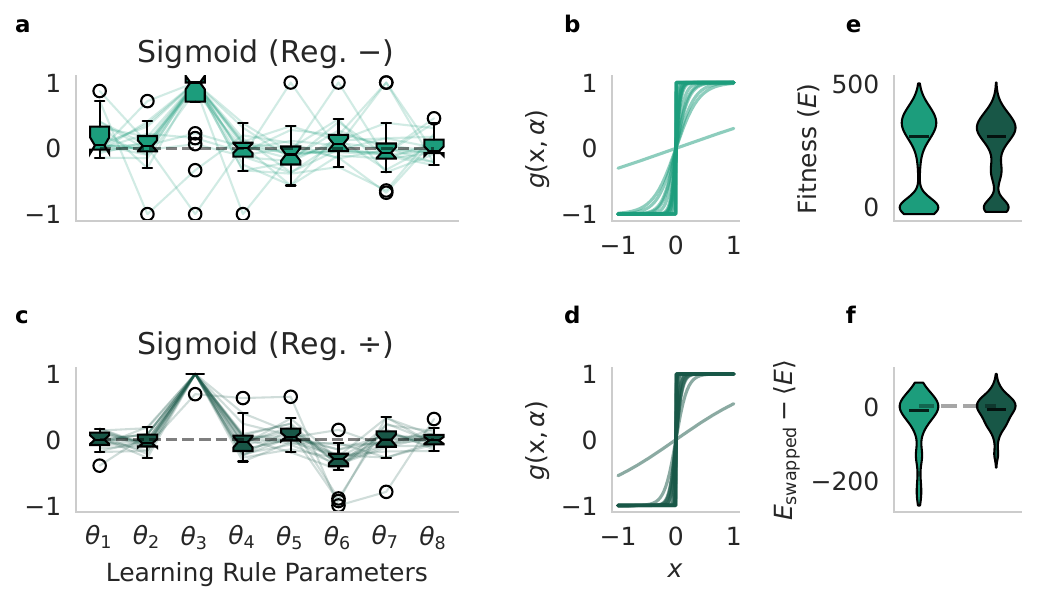}
\caption{\textit{ Trainable readout nonlinearity leads to different rules depending on weight normalization.
\textbf{a}. The evolved plasticity rules across 20 runs for agents with trainable sigmoid sensory network outputs and subtractive weight normalization. \textbf{b}. The sigmoid nonlinearities evolve a range of different slopes for subtractive normalization. \textbf{c}. The evolved plasticity rules with divisive weight normalization are similar to the ones developed for the binary networks with fixed nonlinearity, \fig\ref{fig:Rules_L1}. \textbf{b}. The evolved sigmoid nonlinearities are still broad but steeper in networks trained with divisive normalization. \textbf{e}. The fitness distribution for subtractive (-, light green) vs divisive (÷, dark green) weight normalization. \textbf{f}. The fitness difference between the original and swapped agents.
}}
\label{fig:Rules_Sigmoid}
\end{figure*}

We now assess the impact the nonlinearity of the sensory network has on the evolved learning rule. To test this, we allow the steepness of the nonlinearity to evolve by setting:
\begin{equation}
g(x,  \alpha) = \frac{2}{1 + e^{- \alpha \cdot x}} - 1
\end{equation}
and making $\alpha$ an evolvable parameter. For small values of $\alpha$ the function becomes effectively linear, while for large values $\alpha \to \infty$  it becomes a step function. 

We evolve 40 populations of agents with the trainable sigmoid nonlinearity on the sensory network, half with subtractive and half with divisive weight normalization. For both types of normalization, we observe that in most cases, the resulting learning rules look similar to the ones evolved for the networks with a binary readout (\fig\ref{fig:Rules_Sigmoid}a, b). Still, the two cases are visibly different in terms of the steepness of the evolved sigmoid functions. The divisive normalization evolves higher $\alpha$ (and consequently steeper nonlinearities), while the networks with subtractive normalization evolve relatively steep sigmoids (\fig\ref{fig:Rules_Sigmoid}c), but much more diverse than the ones with divisive normalization (\fig\ref{fig:Rules_Sigmoid}d).

In terms of fitness, both network types perform similarly (\fig\ref{fig:Rules_Sigmoid}e), but when testing agents with motor and sensory networks that did not evolve together (\fig\ref{fig:Rules_Sigmoid}f), the networks with subtractive normalization show a significant ($p_{\text{value}} = 0.00098$) difference in performance, while for the ones with divisive normalization, the performance remains largely unchanged ($p_{\text{value}} = 0.267$). We hypothesize this is due to the motor network adjusting to the specific slope of the sigmoid nonlinearity of its own sensory network (\fig\ref{fig:Rules_Sigmoid}b,d), which makes the more diverse nonlinearities of the networks with subtractive normalization generate some fraction of the swapped agents that cannot perform (fitness close to 0) thus also perform significantly worse on average.

\subsection{Static agents: The distribution of presented foods affects the emerging rule in static networks}

\begin{figure*}[tb]
\centering
\includegraphics[width=0.99\textwidth]{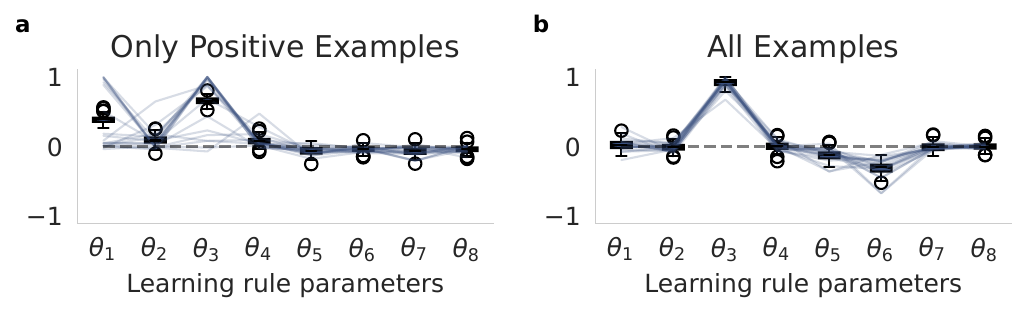}
\caption{\textit{The distribution of presented food values influences the evolved sensory learning rules. The evolved plasticity rules across 20 runs for static sensory networks with binary outputs, trained with divisive normalization. \textbf{a.} The evolved rules when the plasticity step happens only when the network's readout is positive $y_t = 1$. \textbf{b.} The evolved rules when the plasticity step happens after every training step.}}
\label{fig:Rules_Static_Food_Presentation}
\end{figure*}

To better study the convergence of plasticity rules, we remove the motor network and leave only the sensory network subject to the plasticity rule. 
For the following experiments, we define a new \emph{static agent} whose plasticity rule we optimize to perform a simplified version of the task: to decide whether a presented food particle should be consumed or not. We use regularization (from Section~\ref{Sec:Regularization}) and divisive weight normalization to avoid diverging weights. A new food particle is presented to the sensory network each of 125 time steps, and if the output of the network is positive $y_t = 1$ (which we take to mean the agent decides to consume the food particle), feedback is given to the agent and the sensory weights are updated according to the network's learning rule. The network's fitness is assessed by the sum of the values of the foods it consumed. 

After 15 generations, the plasticity parameters converge, at which point the average trained agent at the end of its lifetime can recognize poisonous foods with $ >99 \%$ accuracy. However, the evolved parameters of plasticity rules differ from the ones evolved in the binary-readout moving agents with regularization and divisive normalization  (compare \fig\ref{fig:Rules_Static_Food_Presentation}a and \fig\ref{fig:Rules_L1}f). Specifically, in contrast to the rules from the moving agents where $\theta_3 \rightarrow 1$ and $\theta_1 \rightarrow 0$, in the static networks, we observe that on average $0 < \theta_1, \theta_3 < 1$. We particularly note that in the static networks $\theta_1 + \theta_3 \approx 1$. Moreover, in the static networks, we see that $\theta_6 \rightarrow 0$ in contrast to the moving agent's plasticity rules where on average $\theta_6 < 0$ (\fig\ref{fig:Rules_Static_Food_Presentation}a). We hypothesize that this discrepancy is due to the fact that in the static networks, the plasticity step only occurs when $y_t = 1$, which effectively means the plasticity rule becomes:
\begin{multline}
     \Delta W_t = \eta_p [\theta_1 X_t R_t y_t + \theta_3 X_t R_t] \\ \stackrel{y_t = 1}{=} \eta_p ( \theta_1 + \theta_3) X_t R_t \approx \eta_p X_t R_t
\end{multline}
which is similar to the plasticity rule that evolved in the moving agent's sensory networks, albeit without the negative $\theta_6$ term (\fig\ref{fig:Rules_L1}f).

To test this hypothesis, we repeat the experiment, but now the plasticity step happens at every time step, regardless of the sensory network's output. We now observe that the resulting learning rule is much more similar to the equivalent rule of the moving agents (specifically, we see that $\theta_1 \rightarrow 0, \theta_3 \rightarrow 1 $ and $\theta_1 < 0$ (\fig\ref{fig:Rules_Static_Food_Presentation}b).

This suggests that the original discrepancy between the evolved rules of the static and moving networks was due to the assumption of perfect accuracy in the motor network when simulating only the sensory network (only the food particles to whom the sensory network assigns positive values were consumed). In reality, during a training generation, the moving agents tend to consume quite a few food particles to whom the sensory network assigns a negative value ($y_t = -1$) due to imperfections in the motor network and environmental variability (see the example trajectory, \fig\ref{fig:convergence}a). Thus, the fact that the embodied agents' motor networks make mistakes 
leads to the development of a different learning rule from the one that the agent would evolve in a simplified scenario, where the locomotion of the agent is removed, and the food is instead presented to it sequentially.

\subsection{Different objective functions lead to different evolved rules in static networks}
\begin{figure*}[tb]
\centering
\includegraphics[width=0.99\textwidth]{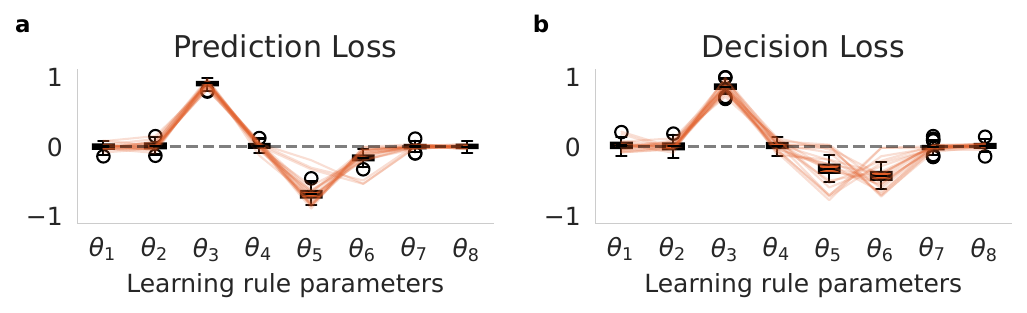}
\caption{\textit{The objective function of sensory learning determines the evolved learning rules. The evolved plasticity rules across 20 runs for static sensory networks with scalar outputs and divisive normalization. \textbf{a.} The evolved rules when the network is optimized to predict the presented food particle's value \textbf{b.} The evolved rules when the network is optimized to decide whether a presented food particle should be consumed or not.}}
\label{fig:Rules_Static_Losses}
\end{figure*}

We finally repeat the same experiment by testing static agents with a linear readout ($g(x) = x$) while following the same normalization standard as before ($L1$ regularization, divisive weight normalization). In this experiment, the readout is a scalar, and thus we train the networks to predict a given food particle's real value as accurately as possible. To do so, we use the mean squared error (MSE) between the actual value of a given food particle and the sensory network's prediction as a training loss for the plasticity parameters $L = \sum_t(R_t - y_t)^2$. 

We see that the resulting rule (\fig\ref{fig:Rules_Static_Losses}a) is not matching the equivalent rule from the moving networks (\fig\ref{fig:Rules_L1}d), but rather evolved the classic Hebbian form $\Delta W_t \approx \eta_p X_t (R_t - y_t) $, associated with subtractive normalization in the moving agents (\fig\ref{fig:Rules_L1}a). 

We hypothesize that the difference in the resulting rules is due to the loss we use to train the static network. In the moving agents, the way the output of the sensory network is processed by the motor network is unclear, but due to the nature of the task, we can relatively safely assume that the only information the motor network really uses is whether a presented food has a positive or negative value. Thus, by testing the static network on the accuracy of its prediction of the presented food's value, we optimized for a different goal than the EA does for the motor networks.

To test this assumption, we retrain the agents with a scalar readout $y_t$, but we assess its performance using the same loss as in the binary networks (if $y_t > 0$, the presented food is consumed, and the agent's total fitness is the sum of all consumed foods). Thus, the plasticity rule has access to a scalar value of the readout (unlike the binary networks, where the $y_t$ that the plasticity sees is always $1$ or $-1$) but is optimized to perform a binary version of the task; decide whether a food is to be consumed rather than predict its value. We find that the resulting rule in this case (\fig\ref{fig:Rules_Static_Losses}b) is more similar to the equivalent rule for the moving agents (\fig\ref{fig:Rules_L1}d), with a strongly negative $\theta_6$ parameter (albeit the static networks also develop a slightly negative $\theta_5$ parameter which is absent for the moving agents' rules). Nevertheless, the similarity suggests that the evolutionary pressure the moving agent's plasticity rule is under is better approximated by the decision rather than the prediction task in the static networks.

\section{Discussion}

We study the emergence of plasticity rules in simple embodied agents that learn to perform a foraging task in a 2D environment. We show how a simple evolutionary algorithm can optimize the different parameters of a linear reward-modulated plasticity rule for solving sensory tasks with sufficient accuracy for guiding the foraging behaviour of embodied agents. Reward-modulated plasticity has been extensively studied as a plausible mechanism for credit assignment in the brain \citep{Razvan2007, Baras2007, Legenstein2008}  and has found several applications in artificial intelligence and robotics tasks \citep{Burms2015, Bing2019}. Here, we demonstrate how such rules can be well-tuned to take into account different parameters and produce optimal behaviour in relatively complex systems with multiple interacting parts.

Recent studies on meta-learning plasticity rules have conclusively shown that identical learning outcomes can be reached via a diverse family of distinct learning rules \citep{Ramesh2023}. Our findings suggest that learning rule redundancy can appear even in very simple systems, such as a single-layer feed-forward linear network. We propose a potential cause of this degeneracy in our network; as the relatively complex motor network is allowed to read out and process the outputs from the plastic network, any consistent information coming out of these outputs can be potentially interpreted in a behaviorally useful way. We find that while introducing an information bottleneck between the sensory and motor networks can enforce some rudimentary structure to the emerging rules, the variability remains high. Still, the introduction of a simple regularization scheme in the evolutionary process \citep{Shervani2023} forces the development of minimal rules that follow interpretable patterns and allow analytical treatment of the learning process. 

Traditionally, plasticity rules are formalized in terms of the pre and post-synaptic activity of individual neurons, despite evidence of heavy bidirectional interaction between synaptic plasticity mechanisms, the structure of individual neurons \citep{Aquin2022}, the topology of neural networks \citep{Butz2014, Giannakakis2023} and various compensatory mechanisms \citep{Zenke2016}. Here, we show that small changes in the neuronal nonlinearity of a single readout unit (linear vs step function) or the weight normalization mechanism (subtractive vs divisive) can strongly affect the evolutionary trajectory of reward-modulated plasticity rules, and lead to convergence on significantly different forms. Additionally, we see that the use of different activation functions and weight normalization schemes can impact the ability of learning rules to generalize across networks. Increased generalizability can potentially enable the mosaic evolution of artificial learning systems, a process that is well documented in biological systems  \citep{Fong2021, Schumacher2022} and is associated with increased specialization and complexity \citep{Aniello2019}. 

We finally study how abstracting the learning process that takes place in the sensory network away from its interaction with the motor network affects the evolved learning rules. We find that small assumptions about important components in the learning process of the moving agents (such as the distribution of values of the food particles the agent consumes) or the precise objective the plasticity is optimizing for (decision vs prediction) can have a strong effect on the resulting plasticity rules. Following the insights of earlier studies \citep{Confavreux2020, Jordan2021, Shervani2023, Mehta2023}, this observation can be used for reverse engineering biologically observed plasticity rules and uncovering their objective function via simulation of simplified (and thus more controlled) learning processes. In particular, comparing the form of a learning rule observed in a complex, embodied system with simplified learning rules, optimized to solve known objective functions, can offer insights into the function of the learning rule in the original system, whose objective function cannot be directly known.

The optimization of functional plasticity in neural networks is a promising research direction both as a means to understand biological learning processes and as a tool for building more autonomous artificial systems. Our results suggest that reward-modulated plasticity is highly adaptable to different environments and can be incorporated into larger systems that solve complex tasks.

 \section{Acknowledgements}
This work was supported by a Sofja Kovalevskaja Award from the Alexander von Humboldt Foundation. EG and SK thank the International Max Planck Research School for Intelligent Systems (IMPRS-IS) for their support. We acknowledge the support from the BMBF through the T\"ubingen AI Center (FKZ: 01IS18039A). AL is a member of the Machine Learning Cluster of Excellence, EXC number 2064/1 – Project number 39072764.

\bibliographystyle{unsrt}
\bibliography{bibliography}

\begin{thebibliography}{10}

\bibitem{Papini2012}
Mauricio~R. Papini.
\newblock {\em Evolution of Learning}, pages 1188--1192.
\newblock Springer US, Boston, MA, 2012.

\bibitem{SnellRood2013}
Emilie~C. Snell-Rood.
\newblock An overview of the evolutionary causes and consequences of behavioural plasticity.
\newblock {\em Animal Behaviour}, 85(5):1004--1011, 2013.
\newblock Special Issue: Behavioural Plasticity and Evolution.

\bibitem{Ellefsen2014}
Kai~Olav Ellefsen.
\newblock The evolution of learning under environmental variability.
\newblock pages 649--656, 07 2014.

\bibitem{Dunlap2016}
Aimee~S Dunlap and David~W Stephens.
\newblock Reliability, uncertainty, and costs in the evolution of animal learning.
\newblock {\em Current Opinion in Behavioral Sciences}, 12:73--79, 2016.
\newblock Behavioral ecology.

\bibitem{Mery2005}
Frederic Mery and Tadeusz~J. Kawecki.
\newblock A cost of long-term memory in <i>drosophila</i>.
\newblock {\em Science}, 308(5725):1148--1148, 2005.

\bibitem{Thornton2011}
Alex Thornton and Tim Clutton-Brock.
\newblock Social learning and the development of individual and group behaviour in mammal societies.
\newblock {\em Philosophical transactions of the Royal Society of London. Series B, Biological sciences}, 366:978--87, 04 2011.

\bibitem{Citri2008}
Ami Citri and Robert Malenka.
\newblock Synaptic plasticity: Multiple forms, functions, and mechanisms.
\newblock {\em Neuropsychopharmacology : official publication of the American College of Neuropsychopharmacology}, 33:18--41, 02 2008.

\bibitem{Feldman2009}
Daniel~E. Feldman.
\newblock Synaptic mechanisms for plasticity in neocortex.
\newblock {\em Annual Review of Neuroscience}, 32(1):33--55, 2009.
\newblock PMID: 19400721.

\bibitem{Wickliffe2019}
Wickliffe Abraham, Owen Jones, and David Glanzman.
\newblock Is plasticity of synapses the mechanism of long-term memory storage?
\newblock {\em npj Science of Learning}, 4, 12 2019.

\bibitem{Power2016}
Jonathan Power and Bradley Schlaggar.
\newblock Neural plasticity across the lifespan.
\newblock {\em Wiley Interdisciplinary Reviews: Developmental Biology}, 6, 11 2016.

\bibitem{he1948organisation}
Donald Hebb.
\newblock The organisation of behaviour, 1948.

\bibitem{BCM1982}
Elie Bienenstock, Leon Cooper, and Paul Munro.
\newblock Theory for the development of neuron selectivity: Orientation specificity and binocular interaction in visual cortex.
\newblock {\em The Journal of neuroscience : the official journal of the Society for Neuroscience}, 2:32--48, 02 1982.

\bibitem{Abarbanel2002}
Henry D.~I. Abarbanel, R.~Huerta, and M.~I. Rabinovich.
\newblock Dynamical model of long-term synaptic plasticity.
\newblock {\em Proceedings of the National Academy of Sciences}, 99(15):10132--10137, 2002.

\bibitem{Hennig2013}
Matthias Hennig.
\newblock Theoretical models of synaptic short term plasticity.
\newblock {\em Frontiers in Computational Neuroscience}, 7, 2013.

\bibitem{Kumarapathirana2021}
K.~P. S.~D. Kumarapathirana, Don Kulasiri, Sandhya Samarasinghe, and Jingyi Liang.
\newblock Computational modelling of synaptic plasticity: A review of models, parameter estimation using deep learning, and stochasticity.
\newblock pages 1--7, 12 2021.

\bibitem{Magee2020}
Jeffrey Magee and Christine Grienberger.
\newblock Synaptic plasticity forms and functions.
\newblock {\em Annual Review of Neuroscience}, 43, 07 2020.

\bibitem{Bengio1991}
Y.~Bengio, S.~Bengio, and J.~Cloutier.
\newblock Learning a synaptic learning rule.
\newblock In {\em IJCNN-91-Seattle International Joint Conference on Neural Networks}, volume~ii, pages 969 vol.2--, 1991.

\bibitem{Soltoggio2018}
Andrea Soltoggio, Kenneth~O. Stanley, and Sebastian Risi.
\newblock Born to learn: The inspiration, progress, and future of evolved plastic artificial neural networks.
\newblock {\em Neural Networks}, 108:48--67, 2018.

\bibitem{Iwhiwhu2020}
Eseoghene Ben-Iwhiwhu, Pawel Ladosz, Jeffery Dick, Wen-Hua Chen, Praveen Pilly, and Andrea Soltoggio.
\newblock Evolving inborn knowledge for fast adaptation in dynamic pomdp problems.
\newblock In {\em Proceedings of the 2020 Genetic and Evolutionary Computation Conference}, GECCO '20, page 280–288, New York, NY, USA, 2020. Association for Computing Machinery.

\bibitem{Najarro2020}
Elias Najarro and Sebastian Risi.
\newblock Meta-learning through hebbian plasticity in random networks.
\newblock In H.~Larochelle, M.~Ranzato, R.~Hadsell, M.F. Balcan, and H.~Lin, editors, {\em Advances in Neural Information Processing Systems}, volume~33, pages 20719--20731. Curran Associates, Inc., 2020.

\bibitem{Pedersen2021}
Joachim~Winther Pedersen and Sebastian Risi.
\newblock Evolving and merging hebbian learning rules: Increasing generalization by decreasing the number of rules.
\newblock In {\em Proceedings of the Genetic and Evolutionary Computation Conference}, GECCO '21, page 892–900, New York, NY, USA, 2021. Association for Computing Machinery.

\bibitem{Yaman2021}
Anil Yaman, Giovanni Iacca, Decebal~Constantin Mocanu, Matt Coler, George Fletcher, and Mykola Pechenizkiy.
\newblock {Evolving Plasticity for Autonomous Learning under Changing Environmental Conditions}.
\newblock {\em Evolutionary Computation}, 29(3):391--414, 09 2021.

\bibitem{Burms2015}
Jeroen Burms, Ken Caluwaerts, and Joni Dambre.
\newblock Reward-modulated hebbian plasticity as leverage for partially embodied control in compliant robotics.
\newblock {\em Frontiers in Neurorobotics}, 9, 2015.

\bibitem{metz2018learning}
Luke Metz, Niru Maheswaranathan, Brian Cheung, and Jascha Sohl-Dickstein.
\newblock Learning to learn without labels, 2018.

\bibitem{lindsey2019learning}
Jack Lindsey.
\newblock Learning to learn with feedback and local plasticity.
\newblock In {\em Real Neurons {\&} Hidden Units: Future directions at the intersection of neuroscience and artificial intelligence @ NeurIPS 2019}, 2019.

\bibitem{Richards2019}
Blake Richards, Timothy Lillicrap, Philippe Beaudoin, Y.~Bengio, Rafal Bogacz, Amelia Christensen, Claudia Clopath, Rui Costa, Archy Berker, Surya Ganguli, Colleen Gillon, Danijar Hafner, Adam Kepecs, Nikolaus Kriegeskorte, Peter Latham, Grace Lindsay, Kenneth Miller, Richard Naud, Christopher Pack, and Konrad Kording.
\newblock A deep learning framework for neuroscience.
\newblock {\em Nature Neuroscience}, 22:1761--1770, 11 2019.

\bibitem{Confavreux2020}
Basile Confavreux, Friedemann Zenke, Everton Agnes, Timothy Lillicrap, and Tim Vogels.
\newblock A meta-learning approach to (re)discover plasticity rules that carve a desired function into a neural network.
\newblock In H.~Larochelle, M.~Ranzato, R.~Hadsell, M.F. Balcan, and H.~Lin, editors, {\em Advances in Neural Information Processing Systems}, volume~33, pages 16398--16408. Curran Associates, Inc., 2020.

\bibitem{Jordan2021}
Jakob Jordan, Maximilian Schmidt, Walter Senn, and Mihai~A Petrovici.
\newblock Evolving interpretable plasticity for spiking networks.
\newblock {\em eLife}, 10:e66273, oct 2021.

\bibitem{Tyulmankov2022}
Danil Tyulmankov, Guangyu~Robert Yang, and L.F. Abbott.
\newblock Meta-learning synaptic plasticity and memory addressing for continual familiarity detection.
\newblock {\em Neuron}, 110(3):544--557.e8, 2022.

\bibitem{Shervani2023}
Navid Shervani-Tabar and Robert Rosenbaum.
\newblock Meta-learning biologically plausible plasticity rules with random feedback pathways.
\newblock {\em Nature Communications}, 14, 03 2023.

\bibitem{confavreux2023metalearning}
Basile Confavreux, Poornima Ramesh, Pedro~J. Goncalves, Jakob~H. Macke, and Tim~P. Vogels.
\newblock Meta-learning families of plasticity rules in recurrent spiking networks using simulation-based inference.
\newblock In {\em Thirty-seventh Conference on Neural Information Processing Systems}, 2023.

\bibitem{Mehta2023}
Yash Mehta, Danil Tyulmankov, Adithya~E. Rajagopalan, Glenn~C. Turner, James~E. Fitzgerald, and Jan Funke.
\newblock Model-based inference of synaptic plasticity rules.
\newblock {\em bioRxiv}, 2023.

\bibitem{Giannakakis2023}
Emmanouil Giannakakis, Oleg Vinogradov, Victor Buendia, and Anna Levina.
\newblock Recurrent connectivity structure controls the emergence of co-tuned excitation and inhibition.
\newblock {\em bioRxiv}, 2023.

\bibitem{Zenke2016}
Friedemann Zenke and Wulfram Gerstner.
\newblock Hebbian plasticity requires compensatory processes on multiple timescales.
\newblock {\em Philosophical Transactions of the Royal Society B: Biological Sciences}, 372(1715):20160259, 2017.

\bibitem{Ramesh2023}
Poornima Ramesh, Basile Confavreux, Pedro~J. Gon{\c c}alves, Tim~P. Vogels, and Jakob~H. Macke.
\newblock Indistinguishable network dynamics can emerge from unalike plasticity rules.
\newblock {\em bioRxiv}, 2023.

\bibitem{Giannakakis2022}
Emmanouil Giannakakis, Sina Khajehabdollahi, and Anna Levina.
\newblock {Environmental variability and network structure determine the optimal plasticity mechanisms in embodied agents}.
\newblock volume ALIFE 2023: Proceedings of the 2023 Artificial Life Conference, 07 2023.

\bibitem{Khajehabdollahi2022}
Sina Khajehabdollahi, Jan Prosi, Emmanouil Giannakakis, Georg Martius, and Anna Levina.
\newblock {When to Be Critical? Performance and Evolvability in Different Regimes of Neural Ising Agents}.
\newblock {\em Artificial Life}, 28(4):458--478, 11 2022.

\bibitem{Aso2014}
Yoshinori Aso, Daisuke Hattori, Yang Yu, Rebecca~M Johnston, Nirmala~A Iyer, Teri-TB Ngo, Heather Dionne, LF~Abbott, Richard Axel, Hiromu Tanimoto, and Gerald~M Rubin.
\newblock The neuronal architecture of the mushroom body provides a logic for associative learning.
\newblock {\em eLife}, 3:e04577, dec 2014.

\bibitem{Fabian2023}
Benjamin Fabian and Silke Sachse.
\newblock Experience-dependent plasticity in the olfactory system of drosophila melanogaster and other insects.
\newblock {\em Frontiers in Cellular Neuroscience}, 17, 2023.

\bibitem{Legenstein2008}
Robert Legenstein, Dejan Pecevski, and Wolfgang Maass.
\newblock A learning theory for reward-modulated spike-timing-dependent plasticity with application to biofeedback.
\newblock {\em PLOS Computational Biology}, 4(10):1--27, 10 2008.

\bibitem{Deb2011}
Kalyanmoy Deb.
\newblock {\em Multi-objective Optimisation Using Evolutionary Algorithms: An Introduction}, pages 3--34.
\newblock Springer London, London, 2011.

\bibitem{Elliott2003}
Terry Elliott.
\newblock {An Analysis of Synaptic Normalization in a General Class of Hebbian Models}.
\newblock {\em Neural Computation}, 15(4):937--963, 04 2003.

\bibitem{Miller1994}
Kenneth~D. Miller and David J.~C. MacKay.
\newblock {The Role of Constraints in Hebbian Learning}.
\newblock {\em Neural Computation}, 6(1):100--126, 01 1994.

\bibitem{Goodhill1994}
Geoffrey~J. Goodhill and Harry~G. Barrow.
\newblock {The Role of Weight Normalization in Competitive Learning}.
\newblock {\em Neural Computation}, 6(2):255--269, 03 1994.

\bibitem{Shen2021}
Yang Shen, Julia Wang, and Saket Navlakha.
\newblock {A Correspondence Between Normalization Strategies in Artificial and Biological Neural Networks}.
\newblock {\em Neural Computation}, 33(12):3179--3203, 11 2021.

\bibitem{Eckmann2022}
Samuel Eckmann and Julijana Gjorgjieva.
\newblock Synapse-type-specific competitive hebbian learning forms functional recurrent networks.
\newblock {\em bioRxiv}, 2022.

\bibitem{Razvan2007}
Răzvan~V. Florian.
\newblock {Reinforcement Learning Through Modulation of Spike-Timing-Dependent Synaptic Plasticity}.
\newblock {\em Neural Computation}, 19(6):1468--1502, 06 2007.

\bibitem{Baras2007}
Dorit Baras and Ron Meir.
\newblock {Reinforcement Learning, Spike-Time-Dependent Plasticity, and the BCM Rule}.
\newblock {\em Neural Computation}, 19(8):2245--2279, 08 2007.

\bibitem{Bing2019}
Zhenshan Bing, Ivan Baumann, Zhuangyi Jiang, Kai Huang, Caixia Cai, and Alois Knoll.
\newblock Supervised learning in snn via reward-modulated spike-timing-dependent plasticity for a target reaching vehicle.
\newblock {\em Frontiers in Neurorobotics}, 13, 2019.

\bibitem{Aquin2022}
Simon d’Aquin, Andras Szonyi, Mathias Mahn, Sabine Krabbe, Jan Gründemann, and Andreas Lüthi.
\newblock Compartmentalized dendritic plasticity during associative learning.
\newblock {\em Science}, 376(6590):eabf7052, 2022.

\bibitem{Butz2014}
Markus Butz, Ines Steenbuck, and Arjen van Ooyen.
\newblock Homeostatic structural plasticity increases the efficiency of small-world networks.
\newblock {\em Frontiers in Synaptic Neuroscience}, 6, 2014.

\bibitem{Fong2021}
Stephanie Fong, Björn Rogell, Mirjam Amcoff, Alexander Kotrschal, Wouter van~der Bijl, Séverine~D. Buechel, and Niclas Kolm.
\newblock Rapid mosaic brain evolution under artificial selection for relative telencephalon size in the guppy (<i>poecilia reticulata</i>).
\newblock {\em Science Advances}, 7(46):eabj4314, 2021.

\bibitem{Schumacher2022}
Erika~L Schumacher and Bruce~A Carlson.
\newblock Convergent mosaic brain evolution is associated with the evolution of novel electrosensory systems in teleost fishes.
\newblock {\em eLife}, 11:e74159, jun 2022.

\bibitem{Aniello2019}
Biagio D'Aniello, Anna Di~Cosmo, Anna Scandurra, and Claudia Pinelli.
\newblock Mosaic and concerted brain evolution: The contribution of microscopic comparative neuroanatomy in lower vertebrates.
\newblock {\em Frontiers in Neuroanatomy}, 13, 2019.

\end{thebibliography}

\end{document}